\documentclass[useAMS, usenatbib]{mn2e}

\usepackage{natbib}

\usepackage{epsfig}
\usepackage{rotating}
\usepackage{graphicx}
\usepackage{epstopdf}
\newif\ifAMStwofonts
\AMStwofontstrue

\title{Submillimetre surveys: The prospects for Herschel}
\author[C. Pearson, S. Khan]
       {Chris Pearson$^{1,2,3}$\thanks{E-mail: Chris Pearson (chris.pearson@stfc.ac.uk) }\thanks{\it{http://www.ir.isas.jaxa.jp/$\sim$cpp/research/}},
Sophia A. Khan$^{4,5}$\\
 $^1$Rutherford Appleton Laboratory, Chilton, Didcot, Oxfordshire OX11 0QX, UK\\
  $^2$Department of Physics, University of Lethbridge, 4401 University Drive, Lethbridge, Alberta T1J 1B1, Canada\\
  $^3$ Department of Physics \& Astronomy, The Open University, U.K. \\
  $^4$Harvard-Smithsonian Center for Astrophysics, 60 Garden Street MS-66 Cambridge, MA 02138, USA\\
  $^5$Shanghai Key Lab for Astrophysics, Shanghai Normal University, Shanghai 200234, China}

\date{Accepted 22nd June 2009.\\
      Received ;\\
      in original form 2009 May }
\pagerange{\pageref{firstpage}--\pageref{lastpage}}

\begin{document}

\label{firstpage}

\maketitle

\begin{abstract}
Using the observed submillimetre source counts, from 250-1200 microns (including the most recent 250, 350 and 500 micron counts from BLAST), we present a model capable of reproducing these results, which is used as a basis to make predictions for upcoming surveys with the SPIRE instrument aboard the Herschel Space Observatory.  The model successfully fits both the integral and differential source counts of submillimetre galaxies in all wavebands, predicting that while ultra-luminous infrared galaxies dominate at the brightest flux densities, the bulk of the infrared background is due to the less luminous infrared galaxy population. The model also predicts confusion limits and contributions to the cosmic infrared background that are consistent with the BLAST results. Applying this to SPIRE gives predicted source confusion limits of 19.4, 20.5 and 16.1mJy in the 250, 350 and 500 micron bands respectively.  This means the SPIRE surveys should achieve sensitivities 1.5 times deeper than BLAST, revealing a fainter population of infrared-luminous galaxies, and detecting approximately 2600, 1300, and 700 sources per square degree in the SPIRE bands (with one in three sources expected to be a high redshift ultra-luminous source at 500 microns). The model number redshift distributions predict a bimodal distribution of local quiescent galaxies and a high redshift peak corresponding to strongly evolving star-forming galaxies.  It suggests the very deepest surveys with Herschel-SPIRE ought to sample the source population responsible for the bulk of the infrared background.

\end{abstract}


\section{Introduction}\label{sec:introduction}

Over two decades ago, a large number of galaxies that emit the bulk of their luminosity in the restframe far-IR were detected in the {\it IRAS} All-Sky Survey (typically at $z<0.1$).  These so-called luminous and ultra-luminous infrared galaxies (LIRGs  $\rm 10^{11}L_{\sun} < L_{IR}$(8-1000$\mu$m)$< 10^{12}L_{\sun}$ and ULIRGs  $\rm L_{IR}\rm >10^{12}L_{\sun}$) are powered by a combination of star formation and active galactic nucleus \citep{soifer87}, but only recently have they been shown to be an important population in the early Universe.  This is in part due to the achievements of submillimetre continuum observations using ground-based facilities: pioneering surveys at 850$\mu$m with SCUBA on the JCMT begat the discovery of submillimetre galaxies (SMGs; e.g., \citealt{smail97},  \citealt{barger98}, \citealt{hugh98}, \citealt{eales99}) which were subsequently constrained to be mainly distant star-forming galaxies (e.g., \citealt{chapman03}).  These characteristics were shared with SMGs found in other submillimetre bands e.g., 1100 \& 1200$\mu$m (\citealt{laurent05}, \citealt{bertoldi00}) and 350\,$\mu$m (\citealt{khan07}).  Larger surveys (e.g. the SCUBA SHADES survey, \citealt{mortier05}) have confirmed these sources are strongly evolving \citep{coppin06}.  However, the discovery of SMGs still poses challenges to semi-analytical hierarchical models of galaxy formation (e.g. \citealt{guiderdoni98}, \citealt{balland03}), and questions remain over their role in the formation of elliptical galaxies and supermassive black holes \citet{magorrian98} and the energy budget between star-formation and accretion in the Universe.

In this work we present a galaxy evolution model that successfully reproduces the source counts from 250-1200$\mu$m, including both the large area SCUBA surveys and the latest results from the BLAST telescope \citep{pascale08}. In Section \ref{sec:model} we describe the model and present fits to the galaxy counts in Section \ref{sec:counts}. The launch of SPIRE on-board the Herschel Space Observatory offers an opportunity to examine an SMG population that overlaps with ground-based observations and IR-luminous galaxies selected at mid -- far-IR wavelengths  (e.g., with IRAS, AKARI, Spitzer).  SPIRE will perform surveys  at 250, 350, 500$\mu$m and in Section \ref{sec:discussion} we discuss the prospects for upcoming surveys with Herschel. Throughout this work a concordance cosmology of  $\rm H_o = 72\,kms^{-1}Mpc^{-1}, \Omega=0.3, \Lambda=0.7$ is assumed.

\section{The Galaxy Evolution Model}\label{sec:model}

To model the submillimetre source counts we use a far-IR backward evolution framework following the models of \citet{cpp05}, \citet{cpp07}. These models were previously successfully used to reproduce the  combined mid-infrared source counts from {\it ISO} \& {\it Spitzer} at 15$\mu$m \& 24$\mu$m. These models have now been updated to produce source counts from 1-1000\,$\mu$m and will be reported in detail in \citet{cpp09}.  
Although submillimetre luminosity functions are available (e.g. \citealt{serjeant05}),  to model the counts we retain the 60$\mu$m luminosity function derived from the {\it IRAS} Point Source Catalogue \citep{saunders00} since it is defined around the peak of the dust emission and contains a large ensemble of sources segregated by population class.  The source counts are fit to the wavelength where the luminosity function is defined, $\lambda _{LF} $, which sets the baseline normalization of all parameters. To predict the counts at other wavelengths, the luminosity function is shifted to the observation wavelength, $\lambda _{obs} $, using the ratio $L(\lambda _{obs})/L(\lambda _{LF} )$, obtained via model template spectra, no other priori is assumed. Spectral templates are drawn from four source populations, comprising normal quiescent galaxies and three star-forming groups consisting of, with increasing luminosity, starburst galaxies ($L_{IR}<10^{11}L_{\sun}$) , LIRGs, and ULIRGs (modelled on the archetype Arp220). An additional AGN component (based on the emission from a dust torus) is also included within the model framework of  \citet{cpp09}, however it is found that AGN do not contribute significantly to the source counts in the submillimetre and although included, their contribution is not considered in this work. The normal galaxy spectral templates are from the libraries described in \citet{efstathiou03} which exhibit cold far-IR/submillimetre colours, with spectra peaking between 100-200$\mu$m. The adopted starburst, LIRG \& ULIRG spectral templates are taken from the spectral models of  \citet{efstathiou00}, which provide good fits to the {\it IRAS}, {\it ISO} and {\it Spitzer} galaxy populations ( \citealt{mrr04},  \citealt{mrr05}).  Note that all templates are independent of the observed data sets being fitted.

\begin{figure}
\centering
\centerline{
\psfig{ figure=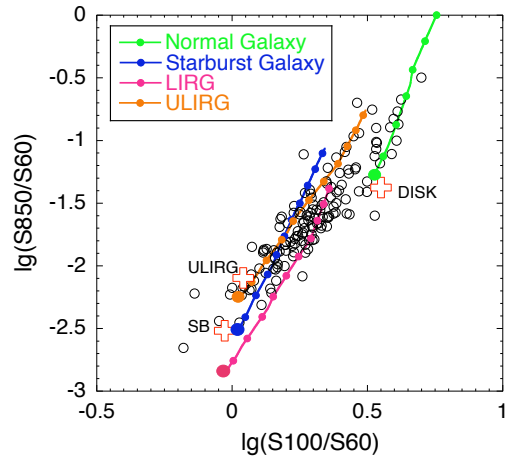,height=6cm}
}
\caption{Colour-colour distributions of the model normal, starburst and U/LIRG templates compared with the local submillimetre \& {\it IRAS} far-infrared colours from Dunne et al. (2000), Vlahakis et al. (2005), with locus of normal (DISK), starburst (SB) and the ULIRG ARP 220. The markers along the SED tracks correspond to redshift steps of $\delta$z=0.1.
\label{sedcolours}}
\end{figure}

Follow-up SCUBA imaging of local IRAS-selected galaxies has implied colder far-IR-submillimetre colours in SMGs than those derived from galaxy spectra based purely on {\it IRAS} colours (\citealt{dunne00},  \citealt{vlahakis05}).  The colours of our model templates agree with this, as they follow the trend of the local galaxy colours extremely well in Figure  \ref{sedcolours}.  Although deeper SCUBA surveys are expected to principally select LIRG/ULIRGs (\citealt{blain02}), this local sample also comprises lower luminosity starburst and cooler normal galaxies (also predicted to contribute at higher redshifts \citealt{efstathiou03}).   

The star-forming populations follow the {\it burst} evolutionary scenario of \citet{cpp05}, \citet{cpp09}, modelled by an exponential function to $z\sim1$ and a power-law thereafter. This evolution is consistent with a rapid onset of star-formation at high redshift, a gradual decline to redshift of $\sim$1 and a sharp decline in activity to the present epoch. The relative contribution of each component to the overall star-formation rate follows a downsizing pattern with redshift in which the most massive galaxies formed stars at an early epoch, thus dominating the star-formation history in the early Universe (e.g.  \citealt{mobasher09}).

%-----------------------------Figure Start------------------------------
%-----------------------------Figure End------------------------------

%-----------------------------Figure Start------------------------------
\begin{figure*}
\centering
\centerline{
\psfig{ figure=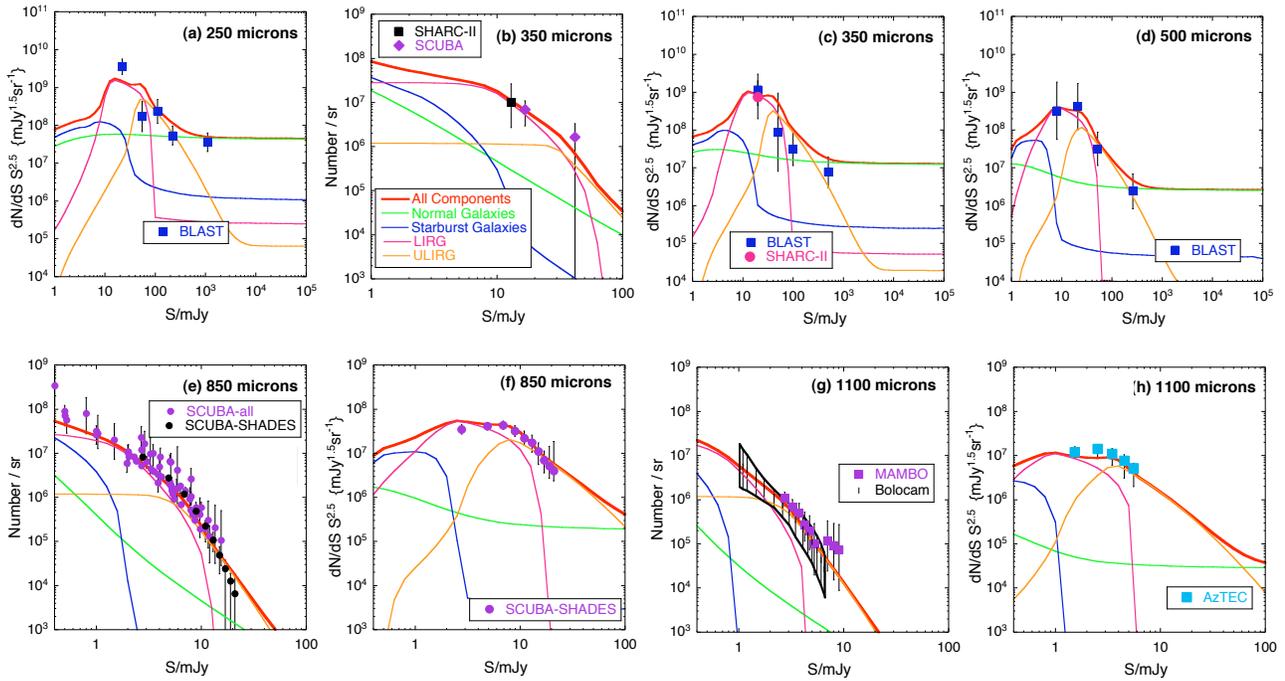,height=9cm}
}
\caption{Model fits to the observed Submillimetre counts from 250 -- 1200$\mu$m.  The total contribution and individual components corresponding to the normal, starburst, LIRG and ULIRG populations are overplotted. {\bf (a,c,d)} Model fits to the observed BLAST counts at 250, 350, 500$\mu$m from Devlin et al. (2009). Also shown is the SHARC-II 350$\mu$m observation from Khan et al. (2007). Source counts are differential normalized to a Euclidean universe. {\bf (b)}The Observed 350$\umu$m integral source counts from the SHARC-II survey of the Bo¬otes field by Khan et al. (2007); also plotted are the normalized 450$\umu$m SCUBA source counts from Smail et al. (2002).  {\bf (e)} Observed 850$\umu$m integral source counts from the various surveys by SCUBA (Smail et al. (1997), Hughes et al. (1998), Eales et al. (1999), Barger et al. (1998), Blain et al. (1999), Smail et al. (2002), Cowie et al. (2002), Scott et al. (2002),  Knudsen et al. (2006)); and the largest SCUBA survey, SHADES, covering $\sim$0.25 sq.deg.  Coppin et al. (2006).  {\bf (f)} The SHADES differential counts normalized to a Euclidean universe.  {\bf (g)} Observed 1100$\umu$m integral source counts are from the  BOLOCAM instrument {\it Black shaded area} from Laurent et al. (2005) and the MAMBO survey of Greve et al. (2004) normalized from 1200$\umu$m.  {\bf (g)} Differential source counts normalized to a Euclidean universe for the AzTEC observations of Perera et al. (2008). 
\label{submmcounts}}
\end{figure*}  
%-----------------------------Figure End------------------------------

\section{Source Counts and Results}\label{sec:counts}

The model fits to the observed source counts at 250, 350, 500, 850 and 1100$\mu$m are shown in Figure \ref{submmcounts}  with the total model source counts shown alongside the respective contributions of the assumed galactic populations (normal and starburst galaxies, LIRGs and ULIRGs).   

Figure \ref{submmcounts} panels {\it a, c} \& {\it d} show the model fits to the differential counts (normalised to a Euclidean universe) from the Balloon-borne Large-Aperture Submillimeter Telescope (BLAST,  \citealt{pascale08}) survey in the GOODS field \citep{devlin09}, for the 250, 350 \& 500$\mu$m bands respectively.  In all the BLAST bands, it is predicted that the brightest counts ($>$1\,Jy in the non-evolving Euclidean regime of the counts) will be dominated by quiescent normal galaxies, expected to be bright, local galaxies at redshifts $<$0.5 (see Figure \ref{nz}). Here the model predicts 2.7 sources at 250 $\mu$m over the BLAST survey area of 8.7 deg$^2$, compared with the three sources found in the brightest bin of the source list of  \citet{devlin09}. The steep departure from Euclidean counts is caused by the ULIRGs but at the peak of the differential source counts the less luminous LIRGs are the dominant population.  At 250\,$\mu$m, the model slightly over-predicts the source counts at $>$60\,mJy, but this is within the BLAST error bars.  There is a sharp rise in the counts at the 200\,mJy level, and a turn-over between 100 and 20\,mJy (although the BLAST counts in this region may be less reliable as the instrument is confusion-limited), with the model predicting a second turn-over at fainter flux densities ($<$10\,mJy).  Due to the strong negative K-corrections in the submillimetre \citep{franceschini91}, the flux densities of distant galaxies are enhanced such that the luminosity function at lower luminosities is sampled at fainter flux densities, with any break in the counts being attributed to a change in the dominant population. 
The BLAST counts are derived from a P(D) analysis rather than source catalogues and provide a statistical constraint on the slope of the source counts at faint fluxes which are already source confused. Encouragingly at 350$\mu$m, the faintest BLAST counts are consistent with the differential counts from the deeper (non-confused) 350\,$\mu$m survey using SHARC~II  in the  Bo¬otes field by \citet{khan07}. Figure \ref{submmcounts}{\it b} shows the 350\,$\mu$m integral source counts from same survey and the SCUBA 450$\mu$m counts (\citealt{smail02}; assuming an Arp 220 spectral template to transform the counts to this band).  The model fits these observations well, predicting breaks in the source counts at $\sim$40 and $\sim$10\,mJy, and that the deeper SHARC~II results are dominated by LIRGs.  In the 500$\mu$m BLAST band, the model fit is exceptionally good, from the steep rise from Euclidean values at $S<$300\,mJy, to the turn-over  between 30-6\,mJy. 
The model predicts turnovers in the counts at fainter flux densities of 10, 8, 5\,mJy in the 250, 350 \& 500$\mu$m bands, all within the constraints imposed by equating the integrated surface brightness of the BLAST sources to the emission from the infrared background derived from a power-law extrapolation and naive cut-off of sources estimated by \citet{devlin09} to be 7.0$\pm$1.3, 7.2$\pm$1.7 \& 4.6$\pm$1.2\,mJy at 250, 350 \& 500$\mu$m respectively.

At longer submillimetre wavelengths, the model fits are compared with the observed integral source counts from the myriad surveys carried out with SCUBA at 850\,$\mu$m (Figure \ref{submmcounts}{\it e}).  These observations span two orders of magnitude in flux density and thus provide the best pre-Herschel constraints on the galaxy counts.  The models provide a good fit to the counts from the brightest flux densities down to 0.5\,mJy -- below the SCUBA-850\,$\mu$m confusion limit of 2mJy (from the lensed surveys of  \citealt{smail97}, \citealt{smail02}). At these levels, due to the strong negative K-corrections, we expect to be able to observe relatively moderate starburst galaxies. At brighter flux densities, $\sim$10\,mJy, ULIRGs are the dominant population (although a significant increasing contribution from normal galaxies cannot be ruled out \citep{efstathiou03}, but from the model they are predicted to dominate at $> \sim$50\,mJy).
The largest 850\,$\mu$m survey to-date (the $\sim$0.25\,deg$^2$ SHADES survey \citealt{mortier05}) detected 120 sources, effectively doubling the number of known SMGs. The SHADES differential source counts \citep{coppin06} are shown in  Figure \ref{submmcounts} {\it (f)}.  The best-fitting model requires a break at $\sim$4-6\,mJy and it is difficult to simultaneously reconcile this with the bright-end counts, using even the most recent evolutionary models (e.g. \citealt{mrr09}). However our model fits both the bright-end counts and this break due to the inclusion of the intermediate LIRG population, which are often omitted in contemporary source count models, between the starburst and ULIRG populations. The break is predicted to be due to the emergence of these strongly evolving galaxies, with their contribution peaking at  $\sim$2\,mJy in the differential counts. 

In Figure \ref{submmcounts}{\it g} the integral source counts at millimetre wavelengths for the surveys with the BOLOCAM instrument at 1100$\mu$m (from the maximum likelihood analysis of \citet{laurent05}) and the MAMBO instrument at 1200$\mu$m (normalising the counts to 1100$\mu$m). BOLOCAM, MAMBO \& SCUBA have all surveyed the same area -- the Lockman Hole -- and in essence, the counts suggest the millimetre observations are sampling the same brighter portion (S$_{850}> \sim$8\,mJy) of the SCUBA 850\,$\mu$m population, expected to be dominated by ULIRGs or even HLIRGs (Hyper Luminous Infra-Red Galaxies, $L_{IR}>10^{13}L_{\sun}$).  This  is simply due to the longer wavelengths sampling further down the Rayleigh-Jeans slope and therefore preferentially selecting  the higher luminosity, high redshift objects.  Finally, the model fits to the differential counts from the recent AzTEC observations of  \citet{perera08} in the GOODS fields are presented in  Figure \ref{submmcounts}{\it h}.  There is a good fit to these observed counts, with the higher luminosity sources providing the main population. The flattening seen in both the integral and differential source counts at fluxes of $\sim$3\,mJy is also reproduced by the model, representing a shift in the dominant population from ULIRGs to LIRGs, with the expectation of further flattening below $\sim$1\,mJy.

\section{The Prospects for Herschel}\label{sec:discussion}

The Herschel Space Observatory \citep{pilbratt08}, launched on 14th May 2009, is ESA's next generation infrared mission. The Spectral and Photometric Imaging Receiver (SPIRE) instrument is one of the focal plane instruments and is designed for photometry and spectroscopy between 200-550$\mu$m  \citep{griffin08}. The three SPIRE bolometer arrays (PSW, PMW and PLW, respectively centered on 250, 350 \& 500$\mu$m, $\lambda / \Delta \lambda \sim$3, with 139, 88, and 43 pixels) allow simultaneous observations over a FOV of 4$\arcmin \times$ 8$\arcmin$ in the three bands. In SPIRE's large map scanning mode the 5$\sigma$, 1 hour point source sensitivities are expected to be 3.7, 5.3 \& 4.6\,mJy for the respective arrays.  Despite being near identical to the three arrays on BLAST, the 3.5m Herschel primary mirror (2m on BLAST) offers superior resolution. 

The ultimate sensitivity of any survey will be the confusion limit, defined as the threshold of fluctuations in the background sky brightness caused by (unresolved) point sources below which sources cannot be discretely detected in the telescope beam $\lambda$/D, where D is the telescope diameter. The confusion due to faint galaxies is more severe at longer wavelengths and smaller apertures and is often characterized by the number of beams per source, with classical limits of 20-40 beams per source often adopted ( \citealt{hogg01},  \citealt{jeong06}). The confusion limits for Herschel-SPIRE and BLAST can therefore be compared using the source count model: for BLAST the 20 beams per source confusion limit is predicted to be 33.7, 33.6 \& 23.9mJy in the 250, 350 \& 500$\mu$m bands -- agreeing very well with the estimates from \citet{devlin09} of 33$\pm$4, 30$\pm$7 \&  27$\pm$4 mJy (implying that the faintest counts reported by BLAST are already source confused). For SPIRE, given the larger aperture, our models predict 20 beams per source confusion limits of  19.4, 20.5 \& 16.1mJy in the 250, 350 \& 500$\mu$m bands respectively, implying that equivalent surveys with Herschel offer a sensitivity improvement of 1.5 over BLAST.  This advantage is significant given the steep nature of the source counts in this flux regime, with the model predicting a shift from the ULIRG dominated counts to LIRG domination. Using  P(D) analysis , the BLAST source counts are able to probe beneath the conventional confusion limit, however P(D) analysis is only able to constrain the slope of the source counts. The intrinsically deeper images afforded by Herschel should be able to resolve and reliably sample this emerging population, which our model predicts will contribute the bulk of the Cosmic Infra-Red Background (CIRB).

%-----------------------------Figure Start------------------------------
\begin{figure}
\centering
\centerline{
\psfig{ figure=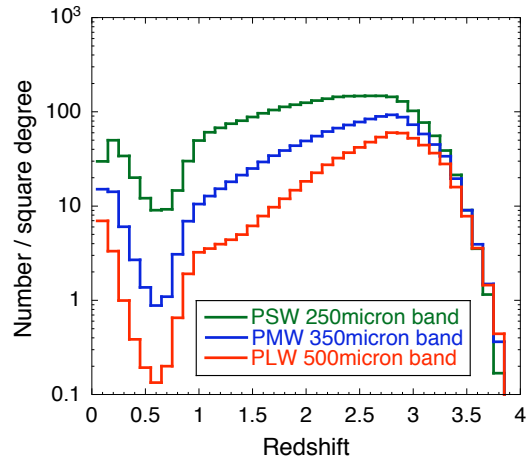,height=6cm}
}
\caption{The number redshift distribution for SPIRE bands at the confusion limit of 19.4, 20.5 \& 16.1mJy for the PSW 250, PMW 350 \& PLW 500$\mu$m arrays. Redshift bin size is $\delta$z = 0.1.
\label{nz}}
\end{figure}  
%-----------------------------Figure End------------------------------

%-----------------------------Figure Start------------------------------
\begin{figure}
\centering
\centerline{
\psfig{ figure=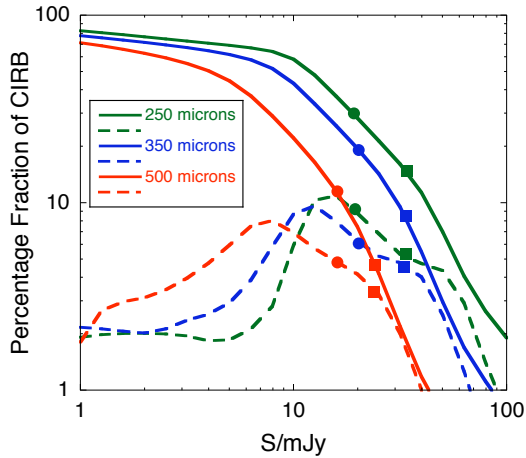,height=6cm}
}
\caption{CIRB fraction as a function of flux for the SPIRE/BLAST bands. {\it solid-lines} are the integral and   {\it dashed-lines} are the differential contribution. Also shown are the confusion limited sensitivities for SPIRE  {\it circles} and BLAST  {\it squares}.
\label{cirb}}
\end{figure}  
%-----------------------------Figure End------------------------------

Using a similar P(D) analysis,  SPIRE surveys could also be pushed well below the corresponding SPIRE confusion limits allowing an actual detection of the turn over in the differential counts  (note that the BLAST survey can only place upper limits). However, surveys with SPIRE to these depths are not expected to resolve the sources responsible for  the break in the source counts (eluded to by  \citealt{khan07} at 350$\mu$m) although this population will be accessible using ground-based facilities such as SHARC-II/CSO, SCUBA-2/JCMT  \citet{holland06}, CCAT  \citet{sebring08}. In the deepest SPIRE confusion-limited surveys, we expect 2600, 1300 and 700 sources per square degree for the 250, 350 \& 500$\mu$m bands respectively, of which $\sim$12$\%$, 25$\%$ \& 35$\%$  will be high redshift ULIRGs. In Figure \ref{nz} we show the number redshift distribution for the three SPIRE bands at a survey sensitivity corresponding to the SPIRE confusion limit. In all bands, but most predominantly in the short wavelength 250$\mu$m band, a bimodal distribution is seen which can be interpreted as a local contribution from quiescent normal galaxies and a high redshift contribution from evolving starburst galaxies, with the high redshift peak becoming more pronounced to longer wavelengths. The median redshift of the N-z distribution lies between $2<z<3$, consistent with the redshift distribution of SCUBA-850\,$\mu$m sources \citep{chapman03}.

Using the model to integrate to the faintest flux levels, the total contribution of faint sources to the cosmic infrared background in the SPIRE bands is estimated, giving intensities of 11.0, 6.0 \& 2.4\,nW/m$^{2}$/sr in the  250, 350 \& 500$\mu$m bands. These values agree well  with the COBE-FIRAS results of 10.4$\pm$2.3, 5.4$\pm$1.6 \& 2.4$\pm$0.6 of \citet{fixen98}. We also estimate that $\sim$80, 85 \& 90$\%$ of the total background resides at z$>$1 at 250, 350 \& 500$\mu$m respectively. In  Figure \ref{cirb} the integral and differential percentage contributions as a function of flux for the SPIRE and BLAST bands are shown, alongside the corresponding 20-beam confusion limits for both instruments.  We expect Herschel to resolve $\sim$30$\%$ (70$\%$), 20$\%$ (60$\%$) \&  12$\%$ (45$\%$) at the confusion (and optimal instrumental 1-hour integration) level in its PSW-250$\mu$m, PMW-350\,$\mu$m \& PLW-500\,$\mu$m arrays, with a corresponding peak in the background emission to occuring at flux densities of 10-25\,mJy, 8-20\,mJy \& 5-10\,mJy in the respective bands.  This implies the very deepest surveys with Herschel-SPIRE should sample the dominant source of the background, which from our model is expected to be mostly LIRGs (rather than more luminous ULIRGs, whose contribution peaks at slightly brighter fluxes in all wavebands).  At faint flux densities ($<$1mJy) 20-30$\%$ of the CIRB remains unresolved and the counts in Figure \ref{submmcounts} indicate that the fainter starburst galaxies will become the dominant population, responsible for the remainder of the total background.

Therefore we expect the upcoming Herschel SPIRE surveys to produce the first large statistically reliable samples of SMGs, taking submillimetre astronomy from the pioneering era into one which detailed constraints can be placed on the evolution of star-formation in the early Universe.

\section{Acknowledgements}
We thank Steve Willner for helpful comments and Andreas Efstathiou for  providing his galaxy templates.  We thank the referee for constructive comments that improved this work.

% *****************************************************************************************
% *****************************************************************************************
% *****************************************************************************************
% *****************************************************************************************

%%%%%%%%%%%%%%%%%%%%%%%%%%%%%%%%%%%%

\label{lastpage}

\end{document}